\documentclass[10pt]{article}
\usepackage{color}
\usepackage{a4wide}
\usepackage{amssymb}
\usepackage{graphicx}
\usepackage{amsmath}
\usepackage{hyperref}
\usepackage{tensor}
\usepackage{slashed}
\usepackage{epstopdf}
\usepackage{extarrows}
\usepackage[utf8]{inputenc}
\def\deltabar{{\mathchar '26\mkern -10mu\delta}}
\newcommand{\td}{\text{d}}

\begin{document}
\title{\bf Comments on Joint Terms in Gravitational Action}

\author{Run-Qiu Yang$^{1}$ and Shan-Ming Ruan$^{2,3}$\\
$^1$\small Quantum Universe Center, Korea Institute for Advanced Study, Seoul 130-722, Korea\\
$^2$\small CAS Key Laboratory of Theoretical Physics, Institute of Theoretical Physics,\\ \small Chinese Academy of Sciences, Beijing 100190, China\\
$^3$\small School of Physical Sciences, University of Chinese Academy of Sciences, No.19A\\ \small Yuquan Road, Beijing 100049, China}

\date{}
\maketitle
%%%%%%%%%%%%%%%%%%%%%%%%%%%%%%%%%%%%%%
\begin{abstract}
This paper compares three different methods about computing joint terms in on-shell action of gravity, which are identifying the joint term by the variational principle in Dirichlet boundary condition, treating the joint term as the limit contribution of smooth boundary and finding the joint term by local SO(1,$d-1$) transformation. In general metric gravitational theory, we show that the differences between these joint terms are some variational invariants under fixed boundary condition. We also give an explicit condition to judge the existence of joint term determined by variational principle and apply it into general relativity as an example.
\end{abstract}

%\end{abstract}
%%%%%%%%%%%%%%%%%%%%%%%%%%%%%%%%%%%%%%

%\tableofcontents
%\flushbottom

%%%%%%%%%%%%%%%%%%%%%%%%%%%%%%%%%%%%%%
\noindent

\section{Introduction}\label{Intro}
Einstein-Hilbert action has been the most simplest action functional for gravity in general relativity. Such action functional (in the case without cosmological constant) is given just by the integration of scalar curvature in a certain space-time region. In this action, the metric and its second derivative are involved, so in principle, the variation problem is well defined only after the metric and its first derivative are both fixed at the boundary. However, motivated by quantum cosmology, Gibbons and Hawking~\cite{PhysRevD.15.2752} had showed that this requirement is too strong. Instead, they suggested that an additional boundary term should be added into the action functional and so that the normal derivative of metric can be cancelled by this boundary term. This boundary term is the Gibbons-Hawking-York boundary term~\cite{PhysRevD.15.2752,PhysRevLett.28.1082}, which is constructed by the trace of extrinsic curvature of the boundary.  With this boundary term, the variation problem is well defined after only the induced metric of the boundary is fixed \cite{Parattu:2015gga,PhysRevD.47.3275}.

The Gibbons-Hawking-York term is defined in the way that it can only be used in non-null surfaces due to the degeneration of induced metric on null hypersurface. Recently, a proposal for the boundary term in null surfaces was given first by Ref.~\cite{Parattu:2015gga} and then carefully discussed in the Refs.~\cite{Parattu:2016trq,Lehner:2016vdi,Hopfmuller:2016scf,Jubb:2016qzt}. Particularly, Refs.~\cite{Parattu:2016trq,Jubb:2016qzt} give two unified boundary terms for null and non-null surface in the coordinates and tetrad form. It is worthy of noting that for null boundary in 4-dimensional space-time, if one fix the conjugate momentum on the null surface, then one does not need to add any boundary term~\cite{Chakraborty:2016yna,Krishnan:2016mcj}. Except for general relativity, the non-null boundary terms for other second order or higher derivative metric theory has also been developed, such as $f(R)$ gravity~\cite{Guarnizo:2010xr,delaCruzDombriz:2009et}, Gauss-Bonnet theory~\cite{0305-4470-14-5-008,PhysRevD.36.392} and Lanczos-lovelock theory~\cite{Padmanabhan:2013xyr,Chakraborty:2017zep}. However, the null boundary terms for these higher order gravity theories are still absent.

Usually, the boundary is assumed to be smooth when we consider what boundary term should be added. It is still worthy of investigating what new things can happen when boundary is piecewise smooth. The space-time boundary with some joints arises naturally in a number of different systems in gravity physics~\cite{PhysRevD.46.1560,Farhi:1989yr} and some theoretical considerations~\cite{PhysRevD.46.620,PhysRevD.47.1407,PhysRevLett.111.261302,PhysRevD.57.2349}. This problem was first studied by Ref.~\cite{PhysRevD.47.3275} in general relativity, which treated the joint intersected by two smooth fragments as the limit of a smooth surface and computed the Gibbons-Hawking-York boundary at this surface. It turns out that this joint has an additional nonzero contribution to the action.

Recently, most motivated by the conjecture named ``complexity-action'' (CA) conjecture~\cite{Brown:2015bva,Brown:2015lvg}, the joint terms in general relativity attract some attentions again.  The CA conjecture emerged from the previous attempts to understand the ER=EPR conjecture~\cite{Maldacena:2013xja,Susskind:2014rva} and states that on-shell action evaluated on a certain subregion of the bulk space-time may be related to the complexity of holographic boundary state.  More exactly, such conjecture says that the complexity of a particular state $|\psi(t_L,t_R)\rangle$ is dual to the on-shell action in the Wheeler-DeWitt (WDW) patch,
\begin{equation}\label{CA-conj}
  \mathcal{C}(|\psi(t_L,t_R)\rangle):=\frac{\mathcal{A}}{\pi\hbar}\,.
\end{equation}
Here $\mathcal{A}$ is the on-shell action of dual gravitational theory in WDW patch. WDW patch is domain of dependence of any Cauchy surface in the bulk whose intersection with the asymptotical boundary are the time slices at $t_L$ and $t_R$.  This conjecture has satisfied several important properties about the complexity. Especially, at the late time limit when $t_R$ or $t_L$ approaches to infinite, it has been shown that it can satisfy complexity growth rate bound,
\begin{equation}\label{dMdtbound}
  \frac{\td\mathcal{C}}{\td t}\leq2M
\end{equation}
in very general cases~\cite{Lloyd2000,Brown:2015lvg,Yang:2016awy}.

%%
%\begin{figure}
%  \centering
%  % Requires \usepackage{graphicx}
%  %\includegraphics[width=0.23\textwidth]{M4.pdf}
%  \includegraphics[width=0.4\textwidth]{SAdS2.pdf}
%  \caption{The Schematic diagram for WDW patch in an eternal asymptotic AdS black hole. For given particular time slices $t_L$ and $t_R$ at the two boundaries, the WDW patch is just the domain of development of any space-like slice which connects $t_L$ and $t_R$, i.e., the yellow region with its boundary. }\label{SAdS}
%\end{figure}
%%

Two obstacles in the application about CA conjecture appear when we try to calculate the on-shell action in the WDW patch. One is that, as the boundary of WDW patch has some null fragments, the CA conjecture has to face this obstacle on computing null boundary terms. The other one is that there are some joints between null boundary and other boundaries, which have some additional contributions to the action. For Einstein's general relativity, the suitable null boundary term was first given by Ref.~\cite{Parattu:2015gga} and then also by Refs.~\cite{Lehner:2016vdi,Jubb:2016qzt} in different methods. See \cite{Carmi:2016wjl,Chapman:2016hwi} for more details about calculations of action in the holographic complexity. The joint terms appearing in the CA conjecture was first given by Ref.~\cite{Lehner:2016vdi}, which determine the joint terms by the requirement that the variational principle in Dirichlet boundary condition(fixing the induced metric at the boundary) should be well defined (we will call the joint term determined by this method as ``variational joint term'' below). However, if the joint connects two space-like or two time-like boundaries, before the CA conjecture, Hayward proposed a method to find the joint term~\cite{PhysRevD.47.3275}, which used a small smooth space-like or time-like surface to replace the joint and identified such joint term as the limit when such small smooth surface approaches to the joint limit.  These two methods are obtained both in the framework of coordinate frame. Ref.~\cite{Jubb:2016qzt} considered the null boundary terms and joints by using Cartan's tetrad formalism where the boundary is only piecewise $C^2$ and the same joint terms also were found from the property of boundary term under local SO(1,3) Lorentz transformation.

Now the thing is interesting. For some cases, we have at least three different methods to identify the joint terms. It has been obviously shown that such three methods can derive the equivalent joint terms in general relativity. But why such three methods can give the same results is not so obvious.  The more important question is that, as it is very interesting to investigate some kinds of higher order gravitational theory in holography and also in CA conjecture~\cite{Cai:2016xho,Alishahiha:2017hwg,Wang:2017uiw,Guo:2017rul}, we should answer if this three methods can give the equivalent joint terms in a general metric gravitational theory. The aim of this paper is to study these three methods and try to find the relationships between them in general  gravitational theory. We will prove that, if the theory is described by an action which is well defined by variational principle, then the differences between these joint terms given by the three methods can only be made from the variation invariants when we fix the same boundary condition and so these three methods are equivalent.
%In addition, we will show that the joint term is necessary to insure that total on-shell action is invariant under diffeomorphism transformation with fixing boundary condition.
We also give an explicit condition to judge the existence of variational joint term and apply it into general relativity as an example.

The organization of this paper is as follows. In section ~\ref{FindJ}, we will first give the exact descriptions about three methods in finding the joint terms and then give the universal proofs on the equivalence between them. In section~\ref{methodvarJ}, we also develop a universal method to judge the existence of variational joint term directly and a procedure to compute joint term. As an example, we apply it into the general relativity and repeat the results obtained by previous references. A brief summary will be found in q	csection~\ref{summ}.

\section{Joint terms from three methods}\label{FindJ}
In this section we will review rapidly the three methods for finding joint terms proposed by Refs.~\cite{PhysRevD.47.3275,Lehner:2016vdi,Jubb:2016qzt} and then prove they are equivalent to each others. Before that, it is worthy of making it clear that what the general metric gravitational theory refers to and some notations we will use in describing the piecewise smooth boundary\footnote{Here the meaning of ``piecewise smooth'' needs to be clarified. In this paper, if the equation of motion or boundary term in action involves derivative of metric or induced metric up to $n$-th order, then  ``piecewise smooth'' boundary means piecewise $C^{n}$ boundary.} in this paper.

We first assume $M$ is a compact $d$-dimensional space-time with boundary $\partial M$. The boundary $\partial M$ is assumed to be piecewise smooth, which is made of a certain of smooth fragments $\Xi_i$ and the joints between these smooth fragments $J_{ij}$. Here we use $J_{ij}$ to stand for the joint between $\Xi_i$ and $\Xi_j$.  In general, it is possible that the joint itself is not smooth, so there are some joints of the joints. These ``joints of the joints'' are the intersections of three and more smooth fragments, which can be written by $J_{ijk}, J_{ijkl},\cdots$. Let $\mathring{\Xi}_i$ stand for the inner region of $\Xi_i$. As the total space-time region $M$ is compact, the boundary of smooth segment $\Xi_j$ are made of some joints and we have $\partial\mathring{\Xi}_j=\{J_{j1}, J_{j2},\cdots, J_{j,j-1},J_{j+1,j},J_{j+2,j},\cdots\}$. We see that $\partial M\neq\bigcup_{i}\mathring{\Xi}_i$, but for any function $f$ which is  bounded in $\bigcup_{i}\mathring{\Xi}_i$ ($f$ may not be bounded  or may not have definition on $\partial\Xi_i$), we have following relationship,
\begin{equation}\label{bounded}
  \int_{\partial M}f\td\mu_{d-1}=\int_{\bigcup_{i}\mathring{\Xi}_i}f\td\mu_{d-1}=\sum_i\int_{\mathring{\Xi}_i}f\td\mu_{d-1}\,,
\end{equation}
where $\td\mu_{d-1}$ is any measurement defined in $\bigcup_{i}\mathring{\Xi}_i$. Specially, we can take $\td\mu_{d-1}$ to be the induced volume element in the smooth fragments. The smooth boundary can be treated as the special case that $J_{ij}=\emptyset$.

Let $I_{\text{bulk}}$ be the bulk action for a general gravitational theory, which is the function of metric $g_{\mu\nu}$ (or the tetrad ${e^I}$ in the Cartan's tetrad formalism) for a given compact space-time region $M$.\footnote{In this paper, Green indices $\mu,\nu,\cdots$ stand for the bulk indices which run the bulk space-time indices. The little Latin indices $a, b$ run over the boundary space-time indices. The capital Latin indices $A,B,\cdots,H$ run over the joint space-time indices. The capital Latin indices $I,J,\cdots$ stands for tetrad indices and run over the whole bulk tetrad indices.} This means we need to add the torsion-free condition for the connection so that all the geometrical quantities are determined by metric or tetrad.  We first need that this bulk action is invariant under the diffeomorphism transformation. General covariant form of action for an arbitrary diffeomorphism invariant theory can be found at \cite{Iyer:1994ys}. Following the guidance and arguments in Refs.~\cite{Parattu:2015gga,Chakraborty:2017zep} that, in a well defined gravitational action, the bulk part itself can determine what boundary terms and joint terms should be added and what should be fixed at the boundary, we see that the variation of the bulk action should have following form~\cite{Chakraborty:2017zep},
\begin{equation}\label{bulkaction1a}
\begin{split}
  \delta I_{\text{bulk}}(M)&=\int_M\td^dx\sqrt{-g}(E_{\mu\nu}\delta g^{\mu\nu} +\nabla_\mu\deltabar v^\mu )\\
  &=\int_M\td^dx\sqrt{-g}E_{\mu\nu}\delta g^{\mu\nu} +\int_{\partial M}d^{d-1}x\deltabar v
\end{split}
\end{equation}
with
\begin{equation}\label{deltav1}
  \deltabar v=\Pi_N\delta Q^N+\delta\mathcal{B}+\deltabar C\,.
\end{equation}
Here $E_{\mu\nu}=0$ just gives the equations of motion. $\Pi_N$ is the canonical conjugate momentum density corresponding to the variable $Q^N$ which should be fixed at the boundary. $\delta B$ is the variation of some geometrical quantities and $\deltabar C$ is extra terms which may cannot be written as the variation of any quantity. Here notation $\deltabar$ stands for it is an infinitesimal value but may not be the variation of any quantity. For example, in general relativity $E_{\mu\nu}$ is just the Einstein tensor and $Q^N$ stand for the induced metric at the boundary if the boundary is non-null(the form of $Q^N$ for null boundary has been given by Ref.~\cite{Parattu:2015gga}). Considering a general covariant gravitational theory with action that is an arbitrary function of $R_{\mu\nu\rho\lambda}$ and $g_{\mu\nu}$, one can find \cite{Padmanabhan:2013xyr}
\begin{equation}
\begin{split}
\deltabar v^c &= 2 \tensor{P}{_a^{bcd}} \delta\tensor{\Gamma}{^a_{bd}} +2\delta g_{bd}\nabla_aP^{abcd}, \\
P^{abcd}&\equiv \frac{\partial L}{\partial \tensor{R}{_{abcd}} },
\end{split}
\end{equation}
which obviously is not easy to be rewritten into the form like (\ref{deltav1}). Some analysis for Lovelovk gravity with non-null boundary recently has been derived in \cite{Chakraborty:2017zep}. In addition, we assume that $Q^N$ in Eq.~\eqref{deltav1} are chosen so that the variational problem is well defined when the boundary is smooth. This assumption means that the  $\deltabar C$ term must be a total derivative term or zero after we fix $\delta Q^N=0$. For the case that $\deltabar C=0$, there is no any difference between smooth and piecewise smooth boundaries. In this paper we will discuss the case that $\deltabar C$ is a total derivative term.

\subsection{Variational joint terms and the equivalence to smooth limit}\label{varjoint}
After we have clarified some fundamental notations and conceptions, we now give the first method to find the joint term, which can be regarded as the generalization of method proposed in Ref.~\cite{Lehner:2016vdi}. In order to make the variational principle well defined for smooth boundary, some suitable boundary terms should be added into bulk action $I_{\text{bulk}}$ to cancel the $\delta B$ term. What's more, we also need to deal with the extra term $\deltabar C$ expected to be expressed as
\begin{equation}\label{deltaA1}
  \deltabar C=\sqrt{-g}\nabla_\mu\deltabar A^\mu,
\end{equation}
where $\deltabar A^\mu$ is an infinitesimal vector field which is tangent to the boundary. Eq.~\eqref{deltaA1} is universal for all kinds of boundaries but for non-null boundary can be simplified as
\begin{equation}\label{deltaA2}
  \deltabar C=\sqrt{|h|}D_a\deltabar A^a,
\end{equation}
 where $h$ is the determinant of induced metric $h_{ab}$ at the boundary and $D_a$ is the covariant derivative determined by induced metric $h_{ab}$.
 Due to the degeneration of $h_{ab}$ on the null surface, we need to introduce another auxiliary null vector to define induced metric and covariant derivative on null boundary. See the appendix in the \cite{Parattu:2015gga} for more discussion.
  %\textcolor{red}{It needs to pay attention that the requirement in the Eqs.~\eqref{deltaA1} and \eqref{deltaA2} means that $\deltabar A^a$ and $\deltabar A$ does not have definitions at the joints.}
If the boundary is smooth, then $\deltabar A^\mu$ is at lest $C^1$. In this case after fixing the variables $Q^N$ at the boundary so that $\delta Q^N=0$, the non-null boundary term in Eq.~\eqref{bulkaction1a} reads,
\begin{equation}\label{bulkaction1}
  \int_{\partial M}\td^{d-1}x(\delta \mathcal{B}+\sqrt{|h|}D_a\deltabar A^a)=\int_{\partial M}\td^{d-1}x\delta \mathcal{B}+\int_{\partial^2 M}\td S_a\deltabar A^a=\int_{\partial M}\td^{d-1}x\delta\mathcal{B}\,
\end{equation}
as $\partial^2M=0$ leads $\int_{\partial^2 M}\td S_a\deltabar A^a=0$. We can also obtain the similar result for null boundary.  This relationship gives the boundary term in general,
\begin{equation}\label{Boundterm1}
  I_{\text{bd}}(\partial M)=-\int_{\partial M}\mathcal{B}\td^{d-1}x\,.
\end{equation}
However, if the boundary is not smooth, after we fix the variables $Q^N$ at the boundary and use the relationship~\eqref{bounded}, the boundary term in Eq.~\eqref{bulkaction1} reads,
\begin{equation}\label{boundaction1}
  \int_{\partial M}\td^{d-1}x(\delta\mathcal{B}+\deltabar C)=\sum_{k}\int_{\mathring{\Xi}_k}\td^{d-1}x(\delta\mathcal{B}+\deltabar C)=-\sum_{k}\delta I_{\text{bd}}(\mathring{\Xi}_k)+\sum_{k}\int_{\mathring{\Xi}_k}\td^{d-1}x \deltabar C\,.
\end{equation}
Here $I_{\text{bd}}(\mathring{\Xi}_k)$ stands for the boundary term~\eqref{Boundterm1} evaluated at the inner region of smooth segment $\Xi_k$. Let $r_a$ be the normal vector of the joint embedded in the smooth fragment and define $\deltabar\tilde{A}=r_a\deltabar A^a\sqrt{\sigma}$ by introducing the induced metric $\sigma_{ab}$ at the joints. By using the Eq.~\eqref{deltaA1}, we reach
\begin{equation}\label{boundaction2}
\begin{split}
  \int_{\partial M}\td^{d-1}x(\delta\mathcal{B}+\deltabar C)&=-\sum_{k}\delta I_{\text{bd}}(\mathring{\Xi}_k)+\sum_{k}\int_{\partial\mathring{\Xi}_k}\td^{d-2}x \deltabar\tilde A\\
  &=-\sum_{k}\delta I_{\text{bd}}(\mathring{\Xi}_k)+\sum_{k>l}\int_{J_{kl}}\td^{d-2}x (-1)^{\alpha_{J_{kl}}}[\deltabar A]_{ij}\,.
  \end{split}
\end{equation}
Here the notation $[\deltabar A]_{ij}$ stands for the difference of limit values of $\deltabar\tilde{A}$ at the two intersectional surface $\Xi_i$ and $\Xi_j$. The $\alpha_{J_{kl}}$ equals to 0 or 1, which depends on the orientation of $J_{ij}$. Now it is clear that the variational principle is well defined if and on if there is geometrical quantity $\eta$ defined on the joint satisfies that,
\begin{equation}\label{Myersjoint1}
  \delta\eta^{\text{variation}}=-(-1)^{\alpha_{J_{kl}}}[\deltabar A]_{ij},~~~\text{when}~\delta Q^N=0\,.
\end{equation}
This gives what joint terms should be added into the total action. In order to distinguish these joint contributions obtained by different methods, we use the notation $\eta^{\text{variation}}$ to show that this joint term is obtained by variational principle. In fact, the Eq.~\eqref{Myersjoint1} cannot determine the joint uniquely. One can add any term which is determined by $Q^N$ into $\eta^{\text{variation}}$ without changing the Eq.~\eqref{Myersjoint1}.

Two assumptions make it possible to define the joint terms. The first one is that the $\deltabar C$ can be expressed as the total divergence term shown in the Eqs.~\eqref{deltaA1} and \eqref{deltaA2} on the smooth fragments after we fix the boundary variables $\delta Q^N=0$. This is necessary even for the smooth boundary and open space-time (without boundary) if we want to obtain a well defined variational problem. The second assumption is the condition that the variation at the joints can be written as the variation term just like what we have shown in the Eq.~\eqref{Myersjoint1}. However, for a particular theory, it is not easy to show whether there is a variable $\eta^{\text{variation}}$ which satisfies the Eqs.~\eqref{Myersjoint1}, as it is easy to compute the variation of a quantity but very difficult to judge if an infinitesimal quantity is a variation of any unknown quantity. In the next part of this subsection and  subsection~\ref{SO13}, we will show that $\eta^{\text{variation}}$ is equivalent to the ones obtained by other two methods. Differing from the implicit definition about the joint term in Eq.~\eqref{Myersjoint1}, the other two methods give the explicit computational approaches to obtain the joint term. As a result, these two methods also give the approaches to check whether the variation problem is well defined when the boundary has some joints.

\begin{figure}
  \centering
  \includegraphics[width=0.41\textwidth]{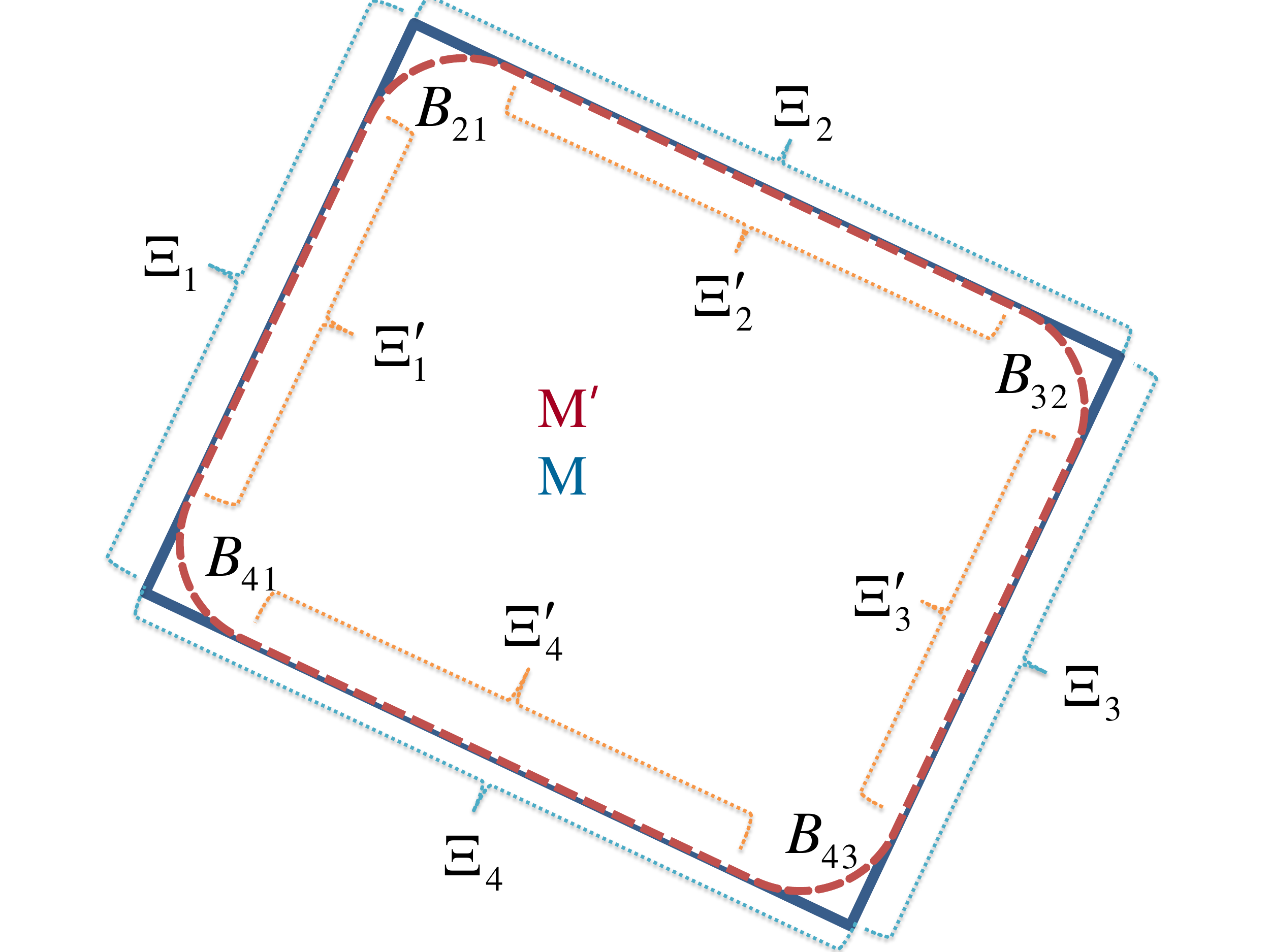}
  \caption{The Schematic diagram for using smooth boundary to replace the joints. At the every joint $J_{ij}$ in the $\partial M$, the joint is replaced by a smooth surface $B_{ij}$ so that $B_{ij}$ is in the inner region of $M$ and can connects the two smooth fragments $\Xi_i$ and $\Xi_j$.}\label{smooth1}
\end{figure}

Let's now consider the method first proposed by Hayward in  Ref.~\cite{PhysRevD.47.3275}. Following Hayward's idea, we now consider a special case that the joint is formed by two smooth fragments which are both space-like or both time-like. In order to find the contributions from the joints, we use some infinitesimal smooth surfaces to replace the joints. One can see the Fig.~\ref{smooth1} as an example. At every joint $J_{ij}$ in the $\partial M$, we use a smooth surface $B_{ij}$ to replace it so that $B_{ij}$ is in the inner region of $M$ and can connect the two smooth fragments $\Xi_i$ and $\Xi_j$ smoothly. The remaining part in every smooth fragment $\Xi_i$ is denoted by $\Xi'_i$. Then the combination of $\{\Xi_i'\}$ and $\{B_{ij}\}$ forms a new smooth closed co-dimensional 1 surface and is the boundary of $M'$. We have,
\begin{equation}\label{boundM2}
  \partial M'=\sum_{k}\Xi'_k+\sum_{i>j}B_{ij}\,.
\end{equation}
It is clear that $M'\subset M$, and in the limit $M'\rightarrow M$, the bulk action defined in $M'$ and $M$ have following relationship,
\begin{equation}\label{IMMprime}
  I_{\text{bulk}}(M)=\lim_{M'\rightarrow M}I_{\text{bulk}}(M')\,,
\end{equation}
and,
\begin{equation}\label{boundXi}
  I_{\text{bd}}(\mathring{\Xi}_k)=\lim_{M'\rightarrow M}I_{\text{bd}}(\mathring{\Xi}_k')\,.
\end{equation}
The joint term in smooth limit then is defined as,
\begin{equation}\label{hayward1}
  \int_{J_{ij}}\td^{d-2}x\eta^{\text{smooth}}:=-\lim_{M'\rightarrow M}\int_{B_{ij}}\td^{d-1}x\mathcal{B}=-\lim_{M'\rightarrow M}\int_{\mathring{B}_{ij}}\td^{d-1}x\mathcal{B}=\lim_{M'\rightarrow M}I_{\text{bd}}(\mathring{B}_{ij})\,.
\end{equation}
Here we use the notation $\eta^{\text{smooth}}$ to stand for joint terms obtained by smooth limit. There are two conditions for the joint terms from  smooth limit. The first one is that the limit in Eq.~\eqref{hayward1} is finite and second one is that, in the limit $B_{ij}\rightarrow J_{ij}$, the boundary integration in the infinitesimal co-dimensional 1 surface $B_{ij}$ can collapse into a integration in the co-dimensional 2 joint $J_{ij}$. Even in the case where the first condition is satisfied, the second one is still a non-trivial requirement to the boundary term. Generally, one may doubt that different choices on the smooth surface may lead different $\eta^{\text{smooth}}$. However, we will see later that if $\eta^{\text{smooth}}$ exists for a particular smooth surface then its value is independent of how to choose the smooth boundaries.

In order to prove the equivalence between the joint term defined by variational principle and smooth limit, let's consider the variation of bulk action in $M'$,
\begin{equation}\label{bulkaction2}
  \delta I_{\text{bulk}}(M')=\int_{M'}\td^dxE_{\mu\nu}\delta g^{\mu\nu}+\int_{\partial {M'}}\td^{d-1}x(\Pi_N\delta Q^N+\delta\mathcal{B}+\deltabar C)\,.
\end{equation}
On the other hand, the Eq.~\eqref{IMMprime} shows that the variation of bulk action in $M$ and $M'$ has also following relationship,
\begin{equation}\label{deltaIMMprime}
  \delta I_{\text{bulk}}(M)=\delta(\lim_{M'\rightarrow M} I_{\text{bulk}}(M'))=\lim_{M'\rightarrow M}\delta I_{\text{bulk}}(M')\,.
\end{equation}
Taking the Eqs.~\eqref{bulkaction2} and \eqref{deltaIMMprime} into account, we can see that,
\begin{equation}\label{boundaction2b}
\begin{split}
  \delta I_{\text{bulk}}(M)&=\lim_{M'\rightarrow M}\left[\int_{M'}\td^dxE_{\mu\nu}\delta g^{\mu\nu}+\int_{\partial M'}\td^{d-1}x(\Pi_N\delta Q^N+\delta\mathcal{B}+\deltabar C)\right]\\
  &=\lim_{M'\rightarrow M}\left[\int_{M'}\td^dxE_{\mu\nu}\delta g^{\mu\nu}+\int_{\partial M'}\td^{d-1}x\Pi_N\delta Q^N+\delta\int_{\partial M'}\td^{d-1}x\mathcal{B}\right]\,.
  %&=\lim_{M'\rightarrow M}\left[\int_{M'}\td^dxE_{\mu\nu}\delta g^{\mu\nu}+\delta\int_{\partial M'}\td^{d-1}x\mathcal{B}\right]\,.
  \end{split}
\end{equation}
Here we have assumed that the Eq.~\eqref{deltaA1} is satisfied so the integration about $\deltabar C$ in the boundary $\partial M'$ is zero. Noting that $\partial M'=(\cup_k\Xi'_k)\cup(\cup_{i>j}B_{ij})$ and the definition about smooth limit joint in Eq.~\eqref{hayward1}, we have,
\begin{equation}\label{boundaction2c}
\begin{split}
  \delta I_{\text{bulk}}(M)&=\lim_{M'\rightarrow M}\left[\int_{M'}\td^dxE_{\mu\nu}\delta g^{\mu\nu}+\int_{\partial M'}\td^{d-1}x\Pi_N\delta Q^N-\sum_{k}\delta I_{\text{bd}}(\mathring{\Xi}'_k)-\sum_{i>j}\delta I_{\text{bd}}(\mathring{B}_{ij})\right]\\
  &=\int_{M}\td^dxE_{\mu\nu}\delta g^{\mu\nu}+\sum_i\left[\int_{\mathring\Xi_i}\td^{d-1}x\Pi_N\delta Q^N-\delta I_{\text{bd}}(\mathring{\Xi}_i)\right]-\lim_{M'\rightarrow M}\sum_{i>j}\delta I_{\text{bd}}(\mathring{B}_{ij})\\
   &=\int_{M}\td^dxE_{\mu\nu}\delta g^{\mu\nu}+\sum_i\left[\int_{\mathring\Xi_i}\td^{d-1}x\Pi_N\delta Q^N-\delta I_{\text{bd}}(\mathring{\Xi}_i)\right]-\sum_{i>j}\int_{J_{ij}}\td^{d-2}x\delta \eta^{\text{smooth}}\,.
  \end{split}
\end{equation}
In this equation, we don't impose the boundary condition $\delta Q^N=0$. On the other hand, by the definition about variational joint term, we have,
\begin{equation}\label{rela1}
  \delta I_{\text{bulk}}(M)\xlongequal[]{\delta Q^N=0}\int_{M}\td^dxE_{\mu\nu}\delta g^{\mu\nu}-\sum_{k}\delta I_{\text{bd}}(\mathring{\Xi}_k)-\sum_{k>l}\int_{J_{kl}}\td^{d-2}x \delta\eta^{\text{variation}}\\
\end{equation}
Now the combination between \eqref{rela1} and \eqref{boundaction2c} immediately implies that,
 \begin{equation}\label{relaeat1}
\delta\eta^{\text{smooth}}\xlongequal[]{\delta Q^N=0} \delta\eta^{\text{variation}}\,.
 \end{equation}
Hence, we see that the variation of two different joint term are the same. This means that the difference between $\eta^{\text{variation}}$ and $\eta^{\text{smooth}}$ can only be made of some zero variational terms when we fix the variables $Q^N$ on the boundary.

In general, we have infinite different choices for $M'$, which leads that there are infinite different choices on surface $B_{ij}$. Let's assume $\eta^{\text{smooth}}_1$ and $\eta^{\text{smooth}}_2$ to be computed by two different kinds of smooth surfaces. As the Eq.~\eqref{boundaction2c} is satisfied for any kind smooth limit, we have
 \begin{equation}\label{relaeat1b}
   \delta\eta^{\text{smooth}}_1=\delta\eta^{\text{smooth}}_2\,,
 \end{equation}
which holds for any kind of variation on the metric. This equation shows that,
 \begin{equation}\label{relaeat1c}
   \eta^{\text{smooth}}_1=\eta^{\text{smooth}}_2+\eta_0\,.
 \end{equation}
Here $\eta_0$ is independent of the metric so  $\eta_0$ is a constant. By the definition, in the limit that the two fragments are connected smoothly, the $\eta^{\text{smooth}}_1$ and $\eta^{\text{smooth}}_2$
should both be zero, so we see that $\eta_0=0$ and $\eta^{\text{smooth}}$ is independent of the choices of smooth connecting surfaces $B_{ij}$.

What's more, we can find from Eq.~\eqref{relaeat1} that if the $\eta^{\text{smooth}}$ exists then it can satisfy Eq.~\eqref{Myersjoint1} and be regarded as the $\eta^{\text{variation}}$. This means that,  a bulk action can give a well defined variational problem by adding some suitable boundary and joint terms if the $\eta^{\text{smooth}}$ exists at every joint which connects two time-like or space-like fragments. As the joint term in smooth limit is defined explicitly in the Eq.~\eqref{hayward1}, this give us a direct method to check whether a bulk action can lead a well defined variation problem in a region with piecewise smooth boundary.

\subsection{SO(1,$d-1$) gauge joint terms and the equivalence to other two methods }\label{SO13}
In this subsection, we will discuss the method proposed by Ref.~\cite{Jubb:2016qzt} to identify the joint term for piecewise smooth boundary. The original framework in  Ref.~\cite{Jubb:2016qzt}  considers variational problem of general relativity in the Cartn's tetrad. They find that the boundary term at the smooth fragments are not invariant under local SO(1,3) transformation which leads an addition contribution at the joint. In this section, we will present the idea in Ref.~\cite{Jubb:2016qzt} in general metric theories and then prove that the joint terms identified by this method are also equivalent to the variational joint terms.

Let us first review the work in Ref.~\cite{Jubb:2016qzt} quickly. Although the conclusions about the joint term in our review are same with that in Ref.~\cite{Jubb:2016qzt}, our explanation about  why we need such joint term is a little different from Ref.~\cite{Jubb:2016qzt}. Ref.~\cite{Jubb:2016qzt} specializes the dimension $d=4$, however, in this paper we will consider general d-dimension. Let $e^I:={e^I}_\mu\td x^\mu$ be the tetrad 1-form. We can define the spin connection 1-form ${\omega^I}_J$ by following Cartan's equation,
\begin{equation}\label{Cartaneq1}
  \td e^I+{\omega^I}_J\wedge e^J=0\,.
\end{equation}
Then the Riemannian curvature tensor 2-form is defined by,
\begin{equation}\label{Riem}
  {\Omega^I}_J:=\td {\omega^I}_J+{\omega^I}_L\wedge {\omega^L}_J\,.
\end{equation}
At the tetrad framework, the Einstein-Hilbert action in the tetrad formulism reads,
\begin{equation}\label{Hilbert1}
  I_{\text{EH}}=\alpha\int_M\varepsilon_{IJ}\wedge \Omega^{IJ} ,
\end{equation}
Where $\Omega^{IJ}:=\eta^{IK}{\Omega^J}_K$, $\eta^{IK}$ is the Minkowski metric, and $(d-r)$-form $\varepsilon_{A_1A_2...A_r}$ is defined as
\begin{equation}
\varepsilon_{A_1A_2...A_r}=\frac{1}{(d-r)!}\varepsilon_{A_1A_2...A_r}e^{A_{r+1}}\wedge ...\wedge e^{A_d}.
\end{equation}
After variation with respective to tetrad $e^I$, one can reach
\begin{equation}
\delta I_{EH}=\alpha \int_{M}[\delta e^K\wedge \varepsilon_{IJK}\wedge \Omega^{IJ} +d(\varepsilon_{IJ}\wedge \delta \omega^{IJ}) ],
\end{equation}
where we have used the torsion-free condition. The first term gives us the equation of motion related to Einstein tensor~\cite{Padmanabhan:2010zzb} and the second term should be cancelled by extra boundary term defined as,\footnote{It is assumed that this boundary term is as same as the boundary term in Eq.~\eqref{Boundterm1}.}
\begin{equation}\label{boundtetrad}
  I_B(\partial M)=-\int_{\partial M}\mathcal{B},~~~\text{with}~\mathcal{B}=\alpha\varepsilon_{IJ}\wedge\omega^{IJ}\,.
\end{equation}
One can see that this boundary term is not invariant under local SO(1,$d-1$) transformation. For a local SO(1,$d-1$) transformation $e^I\rightarrow e'^I={\Lambda^I}_Je^J$, the boundary term will obtain an additional term such that,
\begin{equation}\label{boundtetrad2}
  \mathcal{B}(e'^I)=\mathcal{B}(e^I)-\alpha\varepsilon_{IJ}\wedge(\Lambda^{-1}\td\Lambda)^{IJ}\,.
\end{equation}
Now for a compact space-time $M$ with the piecewise smooth boundary $\partial M$. At the smooth fragments, the boundary term can be directly computed by \eqref{boundtetrad}. However, at the joint, as the the tetrad is discontinuous, the connection 1-form $\omega^{KL}$ has no definition. To overcome this problem, we can use Heaviside function and its derivative $\delta$-function so that the connection $\omega^{KL}$ looks like continuous and we can make integration in Eq.~\eqref{boundtetrad} at the whole boundary. Let $n^\mu$ and $m^\mu$ are the normal vectors for smooth fragments $\Xi_n$ and $\Xi_m$. Define $n^I=n^\mu {e^I}_\mu$ and  $m^I=m^\mu {e^I}_\mu$. For the space-like or time-like normal vector we set $n^In^J\eta_{IJ}=\pm1$ and for the null normal vector we set $n^I\eta_{0I}=1/\sqrt{2}$. The normal vector $m^I$ also obeys the same setting. At the joint $J_{nm}$, the normal vectors at the two sides can be related to each other by a Lorentz transformation,
\begin{equation}\label{Lorentz1}
  n^I={\Lambda^I}_J m^J\,.
\end{equation}
This Lorentz transformation is only defined at the joint $J_{nm}$. For convenience, we will use adapted tetrad on the boundary by requiring that: (1) $e^0$ is the normal vector if the smooth segment is space-like; (2) $e^1$ is the normal vector if the smooth segment is time-like; (3) $(e^0+e^1)/\sqrt{2}$ is the normal vector if the smooth segment is null. After we use adapted tetrad, the Lorentz transformation for the normal vectors is also the Lorentz transformation for the two different kinds tetrad at the joint. However, we need that such transformation should keep the tangent vector space of the joints, so we need  ${\Lambda^I}_J={\delta^I}_J$ when $I,J\neq0,1$.

To extend this SO(1,$d-1$) transformation into a neighborhood in $J_{nm}$, we can introduce a local coordinates $\{s,x^1,x^2,\cdots, x^{d-1}\}$ around joint so that the $x^{d-1}=0$ gives the position of boundary and $x^{d-1}=s=0$ gives the position of joint. We now treat the joint $J_{nm}$ as a kind of ``infinitely narrow codimension 1 surface'' $B_{nm}$, i.e., the surface that $B_{nm}:=\{x^{d-1}=0,s\in(0^-,0^+)\}$. Here $B_{nm}|_{s=0^-}~$ is the limit boundary of $\Xi_n$ and $B_{nm}|_{s=0^+}$is the limit boundary of $\Xi_m$. Then we appoint the tetrad at this ``surface'' as following,
\begin{equation}\label{Lorentz2}
  e^I(s)={\Lambda^I}_Je^J(0^-),~~~\text{with}\quad {\Lambda^I}_J:={\Lambda^I}_J(\eta^{\text{gauge}}\Theta(s))\,.
\end{equation}
Here $\Theta(s)$ is the Heaviside function and $\eta^{\text{gauge}}$ is the rotation/boost parameter between two normal vectors. Eq.~\eqref{Lorentz2} leads that the tetrad looks like ``continuous'' around the joint. After we obtain the tetrad in this ``surface'', we can compute the boundary term in it by the gauge transformation formula in Eq.~\eqref{boundtetrad2}. We can find the that boundary term in this infinitely narrow ``surface'' is,
\begin{equation}\label{gaugebd1}
  I_B(B_{nm})=\int_{J_{nm}}\td^{d-2}x\int_{0^-}^{0^+}\td s\left[-\mathcal{B}|_{s=0^-}+\alpha\varepsilon_{IJ}\wedge(\Lambda^{-1}\td\Lambda)^{IJ}\right]\,.
\end{equation}
As the segment $\Xi_n$ is smooth, $\mathcal{B}|_{s=0^-}$ is finite and so its contribution on the \eqref{gaugebd1} can be removed. The we find the contribution from this infinitely narrow ``surface'' is,
\begin{equation}\label{gaugebd2}
  I_B(B_{nm})=\int_{J_{nm}}\td^{d-2}x\int_{0^-}^{0^+}\td s\alpha\varepsilon_{IJ}\wedge (\Lambda^{-1}\td\Lambda)^{IJ}=\int_{J_{nm}}\td^{d-2}x\eta^{\text{gauge}}\,
\end{equation}
This gives the joint term for general relativity. Here we use the notation $\eta^{\text{gauge}}$ to stand for the joint term obtained by the SO(1,$d-1$) gauge transformation of the boundary term.

The method that we obtain the joint term seems to be different from the one in Ref.~\cite{Jubb:2016qzt} but the Eq.~\eqref{gaugebd2} is just the one obtained by Ref.~\cite{Jubb:2016qzt}. This can be understood physically. When we use a smooth boundary to replace the joint approximately, the contribution on the boundary term will be dominated by the largest component of $\Lambda^{-1}d\Lambda$. This is the same as the gauge transformation in Ref.~\cite{Jubb:2016qzt}. As we zoom in on the joint, connection and tetrad become approximately smooth, and boundary term can give the difference in SO(1,d-1) gauge between the two boundary segments. But there are some new things in our review. Ref.~\cite{Jubb:2016qzt} shows that the gauge transformation of the normal vector between the two sides of the joint can give a finite contribution for the boundary integration. But they did not explain why this term had to be added into the action and why this term was just the contribution of joints. From our process, we can see it clearly that such term is just the contribution of boundary integration at the joint and it has to be added into the total action besides the boundary term at the smooth fragments.

The idea to obtain $\eta^{\text{gauge}}$ can be generalized into other gravitational theory as follows. Firstly, we need the bulk action can be added some suitable boundary for smooth boundary so that the variation problem is well defined. As the tetrad formulism is equivalent to metric formulism, we can convert the variation with respective to metric into the variation with respective to tetrad.

 For every joint $J_{nm}$, assume that the rotation/boost angular between two normal vectors is $\eta_0$. Then we can compute the Lorentz transformation ${\Lambda^I}_{J(nm)}(\eta_0)$ for normal vectors and compute the additional term by this Lorentz transformation just by similar methods in Eqs.~\eqref{Lorentz2} and \eqref{gaugebd2} More detailed, we assume the boundary term $\mathcal{B}$ in the Eq.~\eqref{Boundterm1} can be written in the function of tetrad (with its derivatives) universally for null and non-null boundary fragments and  has following transformation property under any the gauge transformation $e'^I\rightarrow{\Lambda^I}_Je^J$,
\begin{equation}\label{gaugeB1}
  \mathcal{B}(e'^I)=\mathcal{B}(e^I)+\Delta\mathcal{B}(e^I,{\Lambda^I}_{J(nm)})\,.
\end{equation}
Some gravitational theories in tetrad framework tread the spin connection as independent variable so $\mathcal{B}$ would be the functional of spin connection and tetrad. However, in our considerations the spin connection should  also be the function of the tetrad as we have assumed the gravity can be described completely by metric or tetrad. This leads that $\mathcal{B}$ can be determined by $e^I$ (with its derivatives) completely. We choose the adapted tetrad at the boundary then ${\Lambda^I}_J$ gives the Lorentz transformation between two sets of adapted tetrad  at the joint. We introduce a local boundary coordinates $\{s,x^1,x^2,\dots,x^{d-2},x^{d-1}\}$ around joint so that  $x^{d-1}=0$ gives the position of boundary and $x^{d-1}=s=0$ gives the position of joint. Similar to the case Eq.~\eqref{Lorentz2}, we first treat every joint as an infinitely narrow ``surface'' with $x^{d-1}=0$ and $s\in(0^-,0^+)$. Then we extend the gauge transformation at the joint into a neighborhood of $s\in(0^-,0^+)$ and write the rotationa/boost angular for this SO(1,$d-1$) gauge transformation as $\eta(s)=\Theta(s)\eta_0$, by which we can appoint the adapted tetrad in this infinitely narrow ``surface''. If there is a function $\eta^{\text{gauge}}$ so that,
\begin{equation}\label{deltabound1j}
\begin{split}
I_B(B_{nm})&=-\int_{B_{nm}}\td^{d-1}x\mathcal{B}\\
&=-\int_{J_{nm}}\td^{d-2}x\int_{0^-}^{0^+}\td s\mathcal{B}(e^I(s))\\
&=-\int_{J_{nm}}\td^{d-2}x\int_{0^-}^{0^+}\td s\Delta\mathcal{B}[e^I(0^-),{\Lambda^I}_{J(nm)}(\eta(s))]\\
&=\int_{J_{nm}}\td^{d-2}x\eta^{\text{gauge}}
\end{split}
\end{equation}
for any joint $J_{nm}$. Then $\eta^{\text{gauge}}$ is the joint term in SO(1,$d-1$) gauge transformation. Here the integral variables $x^1,x^2,\cdots,x^{d-2}$ run to the whole region of joint and $s\in(0^-,0^+)$. It does not assume that $\eta^{\text{gauge}}$ is equal to $\eta_0$ in general cases (though they are the same in general relativity). We have seen that the $\eta^{\text{gauge}}$ is just the integration of boundary term extended into the joints by SO(1,$d-1$) gauge transformation. It needs to note the condition for Eq.~\eqref{deltabound1j}. For any bulk action, we can always compute the variation of corresponding boundary term under the Lorentz transformation. However, such additional term may not be written into a joint integration. It is a necessary condition for the existence of gauge joint terms that the left hand of Eq.~\eqref{deltabound1j} can be written as an integration in joint.

%One property is that when $\eta^{\text{gauge}}$ is well defined,  the Eq.~\eqref{deltabound1} should hold for any joint, so this equation in fact define function between ${\Lambda^I}_J$ and gauge joint terms. In this sense we can consider the variation of gauge joint term after we make an infinitesimal SO(1,$d-1$) transformation ${\Lambda^I}_{J(nm)}\rightarrow{\Lambda^I}_{J(nm)}+\delta{\Lambda^I}_J$. This infinitesimal transformation will lead the variations of joint term,
%%
%\begin{equation}\label{boundjoint2}
%  \eta^{\text{gauge}}\rightarrow\eta^{\text{gauge}}+\delta_{\Lambda}\eta^{\text{gauge}}\,.
%\end{equation}
%%
%Here we use the index $\Lambda$ in the variation symbol to explicitly show this variation is because of a infinitesimal SO(1,$d-1$) transformation.

Now let's prove that this generalized gauge joint term is equivalent to the othet two joint terms $\eta^{\text{variation}}$ and $\eta^{\text{smooth}}$. For the case the smooth limit can be used, it is obvious that $\eta^{\text{gauge}}=\eta^{\text{smooth}}$ and one of $\eta^{\text{gauge}}$ and $\eta^{\text{smooth}}$ is well defined if and only if the other one is also well defined.\footnote{There we have assumed that the boundary terms obtained by these two methods are the same. As there are some freedom on the choice of boundary term, the boundary terms from these two methods may be different. Then the joint terms $\eta^{\text{gauge}}$ and $\eta^{\text{smooth}}$ can also be different.}  This is because that the infinitely narrow surface $B_{nm}$ is a very special ``smooth surface'' in the sense that we treat Heaviside function is a smooth function. Now let's show that for the general case, the joint terms obtained by gauge transformation and variational method are equivalent to each other.

For an arbitrary variation on the metric with fixed boundary condition $\delta Q^N=0$, we can see that,
\begin{equation}\label{bulkvar1}
  \delta I_{\text{bulk}}=-\int_M\td^dxE^{\mu\nu}\delta g_{\mu\nu}-\sum_k\delta I_B(\mathring{\Xi}_k)-\sum_{n>m}\int_{J_{nm}}\td^{d-2}x\delta\eta^{\text{variation}}\,,
\end{equation}
By the relationship $g_{\mu\nu}=\eta_{IJ}{e^I}_\mu{e^J}_\nu$, we can obtain $\delta g_{\mu\nu}=2\eta_{IJ}{e^I}_{(\mu}\delta{e^J}_{\nu)}$ for any kind of variation of the tetrad. Then we can see that,
\begin{equation}\label{bulkvar1b}
  \delta I_{\text{bulk}}=-2\int_M\td^dx{E^\mu}_I\delta {e^I}_\mu-\sum_k\delta I_B(\mathring{\Xi}_k)-\sum_{n>m}\int_{J_{nm}}\td^{d-2}x\delta\eta^{\text{variation}}\,,
\end{equation}
On the other hand, we can directly compute the variation of bulk action with respective to tetrad ${e^I}_\mu$, which gives,
\begin{equation}\label{bulkvar1c}
  \delta I_{\text{bulk}}=-2\int_M\td^dx{E^\mu}_I\delta {e^I}_\mu-\delta I_B(\partial M)\,.
\end{equation}
Then separating the boundary $\partial M$ into the smooth fragments and joints, we have $I_B(\partial M)=\sum_iI_B(\mathring{\Xi}_k)+\sum_{n>m}I_B(B_{nm})$ and using the definition of $\eta^{\text{gauge}}$ in Eq.~\eqref{deltabound1j}, we find that,
\begin{equation}\label{bulkvar1d}
  \delta I_{\text{bulk}}=-2\int_M\td^dx{E^\mu}_I\delta {e^I}_\mu-\sum_k\delta I_B(\mathring{\Xi}_k)-\sum_{n>m}\int_{J_{nm}}\td^{d-2}x\delta\eta^{\text{gauge}}\,.
\end{equation}
Comparing the Eqs.~\eqref{bulkvar1b} and \eqref{bulkvar1d}, we obtain that,
\begin{equation}\label{relgauge2}
  \delta\eta^{\text{gauge}}\xlongequal[]{\delta Q^N=0}\delta\eta^{\text{variation}}\,.
\end{equation}

\section{Method to find variational joint terms}\label{methodvarJ}
In the section~\ref{varjoint}, we have seen that at the piecewise smooth boundary, the variation problem is well defined if and only if there is term $\eta^{\text{variation}}$ defined in the joint can satisfies the Eq.~\eqref{Myersjoint1}. For a particular bulk action, in principle, there is no difficulty to write its variation into the Eq.~\eqref{bulkaction1} and then find the expression for $[\deltabar A]_{ij}$ and determine the value of $\alpha_{J_{kl}}$ at every joint $J_{kl}$. As we have pointed, it is not a trivial work to verify whether there is any quantity defined in the joint whose variation is just the righthand of Eq~\eqref{Myersjoint1}. Though we have proven that this work can be transformed into finding the smooth limit joint terms and gauge transformation joint terms, it is still very interesting to develop some explicit methods to judge the existence of variational joint term and compute its expression.  We will do these in this section and take general relativity as an example to show how use our method to find out variational joint terms.

\subsection{Condition for variational joint terms}
Before we give out the condition about the existence of variational joint terms, let's first consider an enlightening example in the multi-variable calculus. Let $\vec{x}=(x^1,x^2,\cdots,x^n)$ and $\{f_1(\vec{x}),f_2(\vec{x}),\cdots,f_n(\vec{x})\}$ be a group of functions of $\vec{x}$ which are non-singular in a region $D^n$. Supposing that an infinitesimal quantity $\delta L=f_i(\vec{x})\td x^i$, what is the condition for that there is a function $F(\vec{x})$ such that its differential $\td F=\delta L$? A fundamental theorem in calculation tells us that,
\begin{equation}\label{exisvarJ1}
  \exists F(\vec{x})~\text{such that}~\td F=f_i(\vec{x})\td x^i\Leftrightarrow \frac{\partial f_i}{\partial x^j}-\frac{\partial f_j}{\partial x^i}=0,~\forall \vec{x}\in D^n\,.
\end{equation}
And one expression of $F(\vec{x})$ can be obtained by following single variable integration,
\begin{equation}\label{findF1}
  \forall \vec{x}_2\in D^n,~~~F(\vec{x}_2)=F_1+\int_lf_i\td x^i=F_1+\int_{s_1}^{s_2}f_i[\vec{x}(s)]\frac{\td x^i}{\td s}\td s\,.
\end{equation}
Here $F_1:=F(\vec{x}_1)$ is the value of $F(x)$ at any initial point $\vec{x}_1$ and $l$ is any curve connecting $\vec{x}_1$ and $\vec{x}_2$. The curve $l$ can be parameterized by $\vec{x}=\vec{x}(s)$ and $\vec{x}_1=\vec{x}(s_1), \vec{x}_2=\vec{x}(s_2)$. The integration result is independent of the choices of connecting curve $l$.

In the following, we will show a similar result in the version of variation. Let's assume $\vec{q}(x)=\{q^\alpha(x)\}=\{q^1(x),q^2(x),\cdots,q^n(x)\}$ is a set of independent variation variables. There is a infinitesimal quantity
\begin{equation}\label{integralQ1}
  \deltabar Q=\int_{b}^{a}\td xf_\beta(\vec{q},\vec{q}',\vec{q}'',\cdots,\vec{q}^{(m)})\delta q^\beta\,.
\end{equation}
Here we define $\vec{q}^{(i)}$ is the $i$-th order derivative with respective to $x$, i.e., $\vec{q}^{(i)}=\td^i\vec{q}/(\td x)^i$. $m$ is the highest order of derivative involved in $f_{\beta}$. Now we discrete the space variable $x$ into $x_i=b+(i-1)\Delta x$ with $i=1,2,3,\cdots,\infty$ and assume $\td x\simeq\Delta x=x_{i+1}-x_i$. Define that $q^{i,\alpha}:=q^\alpha|_{x=x_i}$ and $f_{i,\alpha}:=f_\alpha|_{x=x_i}$. All the derivatives of $q^{\alpha}$ can be also written into the discrete forms by using symmetric difference quotient. For example,
\begin{equation}\label{deriv12}
   q'^{i,\alpha}:=q'^{\alpha}|_{x=x_i}=\frac{q^{i+1,\alpha}-q^{i-1,\alpha}}{2\Delta x}\,.
\end{equation}
By this discretion, as $f_\beta$ may contain the derivatives of $\vec{q}$, $f_{i,\beta}$ now becomes the function of $q^{i-m,\alpha}, q^{i-m+1,\alpha},\cdots,q^{i+m,\alpha}$. Then we write this integration \eqref{integralQ1} into the form of infinite summation,
\begin{equation}\label{infinitesum1}
  \deltabar Q\simeq\Delta x f_{i,\alpha}\td q^{i,\alpha}\,
\end{equation}%
with $\alpha=1,2,\cdots,n$ and $i=1,2,3,\cdots,\infty$. The Eq.~\eqref{infinitesum1} can be treated as an $n\times\infty$-variables differential form with independent variables $q^{i,\alpha}$. If we assume the Eq.~\eqref{exisvarJ1} can be generalized into the infinite dimensional case, then we see that there is a quantity $J$ such that $\td J=\deltabar Q$ if and only if,
\begin{equation}\label{exisvarJ2}
  \frac{\partial f_{i,\alpha}}{\partial q^{j,\beta}}-\frac{\partial f_{\beta,j}}{\partial q^{\alpha,i}}=0
\end{equation}
We can recover the this discrete form  in Eq.~\eqref{exisvarJ2} into  the continuous form by using the functional derivative, which reads,\footnote{There we define the functional derivative by $\delta$-function such that $\delta f(q(x))/\delta q(y):=(\partial f/\partial q)|_{x}\delta(x-y)$.}
\begin{equation}\label{exisvarJ4}
\frac{\delta f_{\alpha}(x)}{\delta q^{\beta}(y)}-\frac{\delta f_{\beta}(y)}{\delta q^{\alpha}(x)}=0\,.\\
\end{equation}
However, as a result, the left-hand of Eq.~\eqref{exisvarJ4} is a generalization function(distribution) rather than a number, Eq.~\eqref{exisvarJ4} is lack of proper meaning. Instead, we require that
\begin{equation}\label{exisvarJ}
  \int_b^a\td x\varpi(x)\left[\frac{\delta f_{\alpha}(x)}{\delta q^{\beta}(y)}-\frac{\delta f_{\beta}(y)}{\delta q^{\alpha}(x)}\right]=0
\end{equation}
holds when $y\in(a,b)$ for $\forall \varpi(x)\in\mathfrak{C}$. The auxiliary function set $\mathfrak{C}$ is collection of all functions which are defined in $(a,b)$ and make the integration~\eqref{exisvarJ4} convergent. Then we conclude that there is a quantity $J$ such that $\delta J=\deltabar Q$ if and only if Eq.~\eqref{exisvarJ} holds.

When the condition \eqref{exisvarJ} is satisfied, we can use the similar method in Eq.~\eqref{findF1} to find the expression of $J$. $J$ is the functional of configuration space spanned by $q^\alpha(x)$. Suppose that $r^\alpha(x)$ is any initial point at the configuration space and $J_0=J[r^\alpha(x)]$. In the discrete version, $J$ is the function of $n\times\infty$ variables $q^{i,\alpha}$. For any point $q^{i,\alpha}=p^{i,\alpha}$, we can find a ``curve'' $\gamma$ to connect $q^{i,\alpha}=r^{i,\alpha}$ and $q^{i,\alpha}=p^{i,\alpha}$.
It is more convenient to choose the curve as follows: the first part of $\gamma$ is $\gamma_1$ which connects $(r^{1,\alpha},r^{2,\alpha}, r^{3,\alpha}, \cdots)$ and $(p^{1,\alpha}, r^{2,\alpha}, r^{3,\alpha}, \cdots,)$, the second part is $\gamma_2$ which continues to connect  $(p^{1,\alpha}, r^{2,\alpha}, r^{3,\alpha}, \cdots,)$ and $(p^{1,\alpha}, p^{2,\alpha}, r^{3,\alpha}, \cdots,)$, then third part is $\gamma_3$ which continues to connect  $(p^{1,\alpha}, p^{2,\alpha}, r^{3,\alpha}, \cdots,)$ and $(p^{1,\alpha}, p^{2,\alpha}, p^{3,\alpha}, \cdots,)$ and so on, until we reach the finial point $(p^{1,\alpha}, p^{2,\alpha}, p^{3,\alpha}, \cdots, p^{k,\alpha},\cdots)$. Then we see that,
\begin{equation}\label{infinitesum2}
  J=J_0+\sum_i\int_{\gamma_i}\Delta x f_{i,\alpha}\td q^{i,\alpha}=J_0+\sum_i\int_{s_{i1}}^{s_{i2}}\Delta x f_{i,\alpha}\frac{\td q^{i,\alpha}}{\td s}\td s\,
\end{equation}
Here we parametrize every $\gamma_i$ by $s$ so that $s_{i1}$  and $s_{i2}$ correspond to the starting and ending points of $\gamma_i$. After transforming Eq.~\eqref{infinitesum2} into the continuous version, we have,
\begin{equation}\label{infinitesum3}
  J[\vec{p}]=J_0+\int_a^b \td x\int_{s_1(x)}^{s_2(x)}f_{\alpha}\frac{\td q^{\alpha}(s,x)}{\td s}\td s\,
\end{equation}
Although in our proof $\vec{q}$ are single variable functions, it is not difficult to generalize Eqs.~\eqref{exisvarJ} and \eqref{infinitesum3} into the case that $\vec{q}$ are multiple variables functions.

\subsection{Example: joint terms in general relativity}
In this subsection we will apply Eqs.~\eqref{exisvarJ} and \eqref{infinitesum3} into general relativity and find the variational joint term. Though Refs.~\cite{Lehner:2016vdi} has given the variational joint term, it dose not give out a universal method to judge existence of the joint and find out its expression. We will show how to use Eqs.~\eqref{exisvarJ} and \eqref{infinitesum3} to deal with this problem in general relativity. One can see this method is also suitable for other gravitational theory.

Let's first quickly review how to find the righthand in Eq.~\eqref{Myersjoint1} and then prove it is a total variation term.  We will follow the notations in Refs.~\cite{Lehner:2016vdi}. For convenience, let's $\{x^\mu\}$ is the coordinates of the bulk space-time $M$ and assume the boundary $\partial M$ is described by a scalar field $\Phi(x^\alpha)=0$. We can introduce the local coordinates $\{y^a\}$ in the boundary. As the boundary is covered by both coordinates $\{x^\mu\}$ and $\{y^a\}$, this induced a map from $\partial M$ to $V$ by $x^\mu=x^\mu(y^a)$. Following the Ref.~\cite{Lehner:2016vdi}, we define the bull back map as,
\begin{equation}\label{fRpullback}
  {e^\alpha}_{a}:=\frac{\partial x^\alpha}{\partial y^a}\,,
\end{equation}
which can pull back any bulk covariant tensor field in the boundary to boundary covariant tensor field. For example, the induced metric field of the boundary is the pull back of bulk metric in the boundary,
\begin{equation}\label{inducedmetric}
  h_{ab}:={e^\alpha}_{a}{e^\beta}_{b}g_{\alpha\beta}|_{\partial M}\,.
\end{equation}
We also define ${e^a}_\alpha=h^{ab}g_{\alpha\beta}{e^\beta}_b$. Under the metric variation, the variation of ${e^\alpha}_{a}$ is zero but the variation of ${e^a}_\alpha$ is not zero in general. In the non-null case, we can define the unit normal vector $n^\mu$ and the induced metric $h_{ab}$ for the boundary $\partial V$. Ref.~\cite{Lehner:2016vdi} has showed that and variation of Einstein-Hilbert action can be written as into Eq.~\eqref{bulkaction1a} and,
\begin{equation}\label{fRvara}
  \deltabar v=\sqrt{|h|}(-2\delta K-\varepsilon D_a\deltabar A^a+K_{ab}\delta h^{ab})\,.
\end{equation}
Here $K_{ab}$ is the extrinsic curvature of $\partial M$, $\varepsilon=1$ if the boundary is time-like and $-1$ if the boundary is space-like, $\deltabar A^a=-\varepsilon{e^a}_\alpha\delta n^\alpha=\varepsilon n^\alpha\delta{e^a}_\alpha$ with the normal vector $n^\mu$ at the boundary $\partial M$.

\begin{figure}
  \centering
  \includegraphics[width=0.6\textwidth]{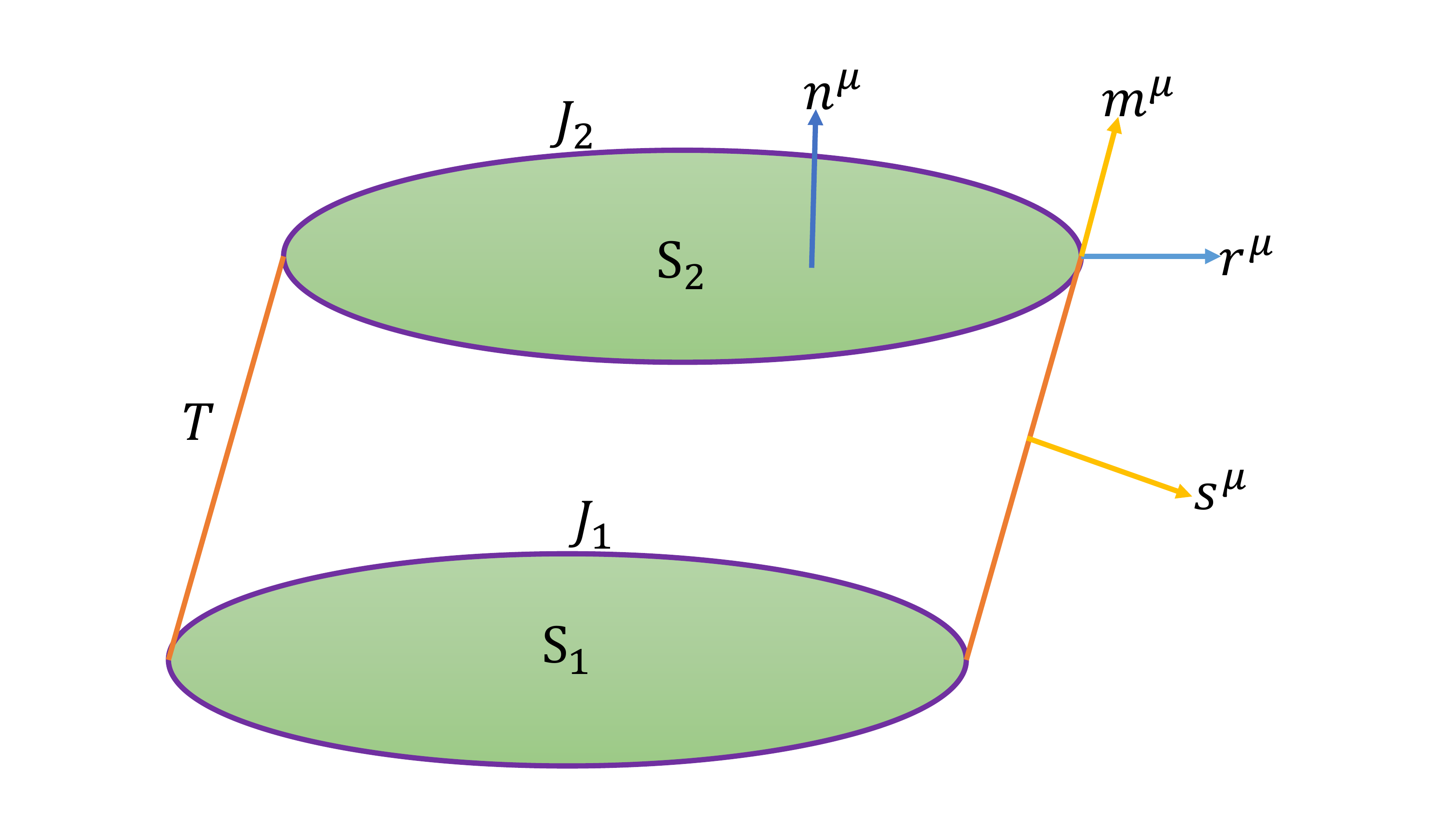}\\
  \caption{The space-time region $M$ bounded by a closed surface $\partial M$ which contains two space-like surfaces $B_1$ and $B_2$ and a time-like surface $T$. There are two joints $J_1$ and $J_2$. }\label{GRjoint1b}
\end{figure}

Let's use a space-like joint $J_2$ as an example. One can see the Fig.~\ref{GRjoint1b}, a space-time region $M$ has a closed boundary $\partial M=S_1\cup S_2\cup T$. Here $S_1$ and $S_2$ are space-like fragments and $T$ is time-like fragment. The space-like joint $J_2$ is the intersection of $S_2$ and $T$. Assuming $n^\mu$ to be the normal vector of $S_2$ and $r^\mu$ is the normal vector of $J_2$ embedded in $S_2$, $s^\mu$ is the normal vector of $T$ and $m^\mu$ is the normal vector of $J_2$ embedded in $T$. Ref.~\cite{Lehner:2016vdi} has shown that the bulk metric variation will lead an additional infinitesimal quantity at the $J_2$,
\begin{equation}\label{GRjoint2}
  \deltabar Q=-\int_{J_2}\td^2x\sqrt{\sigma}(r^\mu n^\nu+m^\mu s^\nu)\delta g_{\mu\nu}
\end{equation}
Here $\sigma=$det[$\sigma_{AB}$] and $\sigma_{AB}$ is the induced metric in $J_2$.

Now let's first directly show that the righthand of Eq.~\eqref{GRjoint2} is a total variation term. As the vector pairs $(n^\mu, r^\mu)$ and $(m^\mu, s^\mu)$ can both form the complete basic for the 2-dimensional vector space normal to $J_2$, we can write,
\begin{equation}\label{nrms1}
  n^\mu=\cosh\eta~m^\mu+\sinh\eta~s^\mu, r^\mu=\cosh\eta ~s^\mu+\sinh\eta ~m^\mu\,
\end{equation}
for a suitable parameter $\eta$. In general $\eta=\eta(x^A)$, which may not be a constant along the joint. As at the boundary, the induced metric $h_{ab}$ and bull back map ${e^\mu}_a$ have been fixed, the relationship $g^{\mu\nu}=-n^\mu n^\nu+{e^\mu}_a{e^\nu}_bh^{ab}$ shows that the real independent variation variables is $n^\mu$ and we can see that $\delta g^{\mu\nu}=-\delta n^\mu n^\nu-n^\mu \delta n^\nu$.\footnote{This is not the unique choice. Alternatively, we can use $s^\mu$ as the independent variation variables.} We can rewrite Eq.~\eqref{GRjoint2} into following form by Eq.~\eqref{nrms1},
\begin{equation}\label{GRjoint3}
  \deltabar Q=-2\int_{J_2}\td^2x\sqrt{\sigma}\cosh\eta~s^\alpha g_{\alpha\beta}\delta n^\beta=-2\int_{J_2}\td^2xf_\beta\delta n^\beta
\end{equation}
Here we define $f_\beta:=\sqrt{\sigma}\cosh\eta~s^\alpha g_{\alpha\beta}$.

Keep in the mind that the boundary geometry and the coordinates are fixed so the vector $r^\mu$ and $m^\mu$ is invariant under variation. Then we can find following useful rules when we change $n^\mu$,
\begin{equation}\label{GRvarrules}
  \frac{\partial g_{\alpha\beta}}{\partial n^\gamma}=n_\alpha g_{\beta\gamma}+n_\beta g_{\alpha\gamma},~\frac{\partial s^\alpha}{\partial n^\beta}=(s^\alpha s_\beta-{\delta^\alpha}_\beta)\sinh\eta-n^\alpha s_\beta,~~\frac{\partial\eta}{\partial n^\beta}=s_\beta\cosh\eta\,.
\end{equation}
By these partial derivative relationships, one can easy check that $\forall \varpi\in\mathfrak{C}$,
\begin{equation}\label{exisvarJ5}
\begin{split}
  &\frac{\delta f_{\alpha}(x^A)}{\delta n^\beta(y^A)}=-2\delta(x^A-y^A)\sqrt{\sigma(x^A)}\sinh2\eta(x^A)~s_\alpha(x^A) s_\beta(x^A)\\
  \Rightarrow&\int\td^{d-2}x\varpi(x^A)\left[\frac{\delta f_{\alpha}(x^A)}{\delta n^\beta(y^A)}-\frac{\delta f_{\beta}(y^A)}{\delta n^{\alpha}(x^A)}\right]=-2\varpi\sqrt{\sigma}\sinh2\eta~(s_\alpha s_\beta-s_\beta s_\alpha)|_{y^A}=0\,.
  \end{split}
\end{equation}
Hence, the $\deltabar Q$ must be a total variation term. Once we obtain the proof for existence, we can use Eq.~\eqref{infinitesum3} to find the corresponding joint term,
\begin{equation}\label{GRjoint1}
  J=J_0-2\int_{J_2}\td^2x\sqrt{\sigma}\int_{\tau_1}^{\tau_2}\cosh\eta~s^\alpha g_{\alpha\beta}\frac{\td n^\beta}{\td \tau}\td\tau\,.
\end{equation}
Here $n^\mu=n^\mu(\tau)$ is any kind of parameterization. One method is that we use the boost angular $\eta$ to parameterize $n^\mu$.
Noting that $r^\mu$ and $s^\mu$ are fixed, we see that, one can use the result in Eq.~\eqref{GRvarrules} to find,
\begin{equation}\label{GRjoint1}
  J=-2\int_{J_2}\td^2x\sqrt{\sigma}\int_{0}^{\eta}\cosh\eta~s_\beta \frac{\td n^\beta}{\td \eta}\td\eta=-2\int_{J_2}\td^2x\sqrt{\sigma}\eta\,.
\end{equation}
This is just the result shown in Ref.~\cite{PhysRevD.47.3275} and then discovered again in Ref.~\cite{Lehner:2016vdi}.

Let's make a brief summary on what we have done for obtaining the Eq.~\eqref{GRjoint1}. First we should determine the real variation variables at the boundary and write the infinitesimal quantity at the joint into the combination of real variation variables. Then we need to use  Eq.~\eqref{exisvarJ} to check if such term is a total variation term. If it does not satisfy the Eq.~\eqref{exisvarJ}, then the variation problem is not well defined at the piecewise smooth boundary. If it satisfies the Eq.~\eqref{exisvarJ}, then we can use Eq.~\eqref{infinitesum3} to find the variational joint term. Of course, in general relativity, if one note the last term in Eq.~\eqref{GRvarrules} and compare it with Eq.~\eqref{GRjoint3}, one can immediately find the result in Eq.~\eqref{GRjoint1}. However, this is just a coincidence in general relativity. Our method is universal and gives a explicit method to judge the existence of variational joint term and find out it if it exists.

\section{Summary}\label{summ}
In this paper, we have compared three different methods about computing joint term in gravitational action, which are identifying the joint term by the variational principle in Dirichlet boundary condition, treating the joint term as the limit contribution of smooth boundary and finding the joint term by local SO(1,$d-1$) transformation. In general metric gravitational theory, if the theory has a well defined variational problem in smooth boundary case,  we have shown that the differences between the joint terms given by these three methods can only be made from some variation invariants of the boundaries and so these three methods are equivalent.

Especially, our proof shows that we can use the smooth limit procedure or local SO(1,$d-1$) transformation to verify whether the variation problem is well defined when some joints appear in the boundary.  For the gauge transformation joint term, we make an explanation on why we need to add this term into the action. Our result shows that the $\eta^{\text{gauge}}$ is just the integration of boundary term extended into the joints by SO(1,$d-1$) gauge transformation. We also develop a necessary and sufficient condition to judge the existence of variational join term directly and a procedure to compute variational joint term. As an example, we apply it into the general relativity and repeat the results obtained by previous references.

It needs to emphasize that the variational joint term can be different from the other two. We say the variational joint term is equivalent to the others, which just means that Eqs.~\eqref{relaeat1} and \eqref{relgauge2} are correct. If one realize that the variational joint term itself is not unique and difference of any two variational joint terms is a variational invariant under fixing $\delta Q^N$, then he can find that Eqs.~\eqref{relaeat1} and \eqref{relgauge2} are the highest conclusions on the relationship between variational joint term and the other two. In fact, the variational joint term depends on the boundary condition. Fixing different boundary condition may leads different variational joint terms and even the boundary terms. For example, in general relativity, if one fix the metric at the boundary, then in principle we do not need to add joint terms at the joints or take $\eta^{\text{variation}}=0$. However, one can easily see that Eqs.~\eqref{relaeat1} and \eqref{relgauge2} are still correct for this kind of boundary condition. The smooth limit joint term and SO(1,$d-1$) joint term depend on which boundary term is used and we have freedom in choosing boundary term, so in principle, one may obtain three different joint terms. However, our proofs show that the Eqs.~\eqref{relaeat1} and \eqref{relgauge2} are still correct if we fix the same boundary conditions during the variations. As in Ref\cite{PhysRevD.47.3275}, the integrand in joint term can be considered as dihedral angle of the joint when the signature of spacetime is Euclidean. But when we move to spacetime with Lorentzian signature, joint term may acquire imaginary piece \cite{Neiman:2013lxa,PhysRevLett.111.261302}which is related to black hole entropy\cite{Neiman:2013ap}.
%It also needs to note that the boundary and joint terms defined by the variational problem may lead the total action (on-shell or off-shell) is not invariant under general diffeomorphsim transformation without fixing $Q^N$ at the boundary. This is because that there is no direct casual relation  between that  variational problems is well defined and the total action is diffeomorphsim invariant.

\section*{Acknowledgments}
We would like to thank Li-Ming Cao for the useful discussions with him. We also would like
to thank ‘The 2017 Annual Meeting of Gravitation and Relativistic Astrophysics and The Fifth
Galileo-Xu Guangqi Meeting’ in Chengdu, China, where part of this work was carried out
during this meeting.
%==============================================================================
\bibliographystyle{unsrt}
\bibliography{ref}

\end{document}